\documentclass[onecolumn,amsmath,amssymb]{snp}
\usepackage{graphicx}
\usepackage{dcolumn}
\usepackage{bm}
\usepackage{indentfirst}  
\usepackage{bm}    
\usepackage{graphicx} 

\def\thefootnote{} 

\renewcommand{\thetable}{\arabic{table}}
\begin{document}
\pagenumbering{arabic}
\title{{\Large Mass spectra of dimesonic states in light flavour sector}}
\author{\large D P Rathaud }
\email{dharmeshphy@gmail.com; raiajayk@gamil.com 
 }
\author{\large A K Rai}
\affiliation{Department of Applied Physics, Sardar Vallabhbhai National Institute of Technology, Surat - 395007, Gujarat, INDIA}

\begin{abstract}
Masses of the dimesonic (meson-antimeson) molecular states are calculated using Yukawa like and meson exchange potential in semi-relativistic approach. The digamma decay width of the dimesonic systems are evaluated using the spectroscopic parameters. Many states such as  $f{}_{0}$(980),   $h{}_{1}$(1380), $a{}_{0}$(1450), $f{}_{0}$(1500), $f{}_{2}'$(1525), $f{}_{2}$(1565), $a{}_{2}$(1700),  $f{}_{0}$(1710),  $f{}_{2}$(1810), $f{}_{2}$(2010) have shown the signature of  dimesonic molecules. Their masses and digamma widths are in good agreement with other theoretical results as well as experimental measurements.\\



\end{abstract}

\maketitle

\section{Introduction}

The exotic states are the states which could not be explained by conventional mesonic ($q\overline{q}$) and baryonic ($qqq$) scheme. Nowadays, the experimental facilities across the world, generating the huge amount of data (states) which need extra theoretical attention \cite{21,1,2,3,4}. All these states carried the information of the fundamental behaviour of the QCD. As per the underlying theory, only the color neutral states are possible, just like mesons and baryons. The theory predicted more complicated color neutral states like tetraquark, pentaquark, hexaquark, hybrid, glueball and molecule like structures \cite{5,Weinstein,6,Weinstein2,7}. The explanation of structures and dynamics of such states are still open questions.  \\

Recently, some experimental groups like, BES III-collaboration \cite{M. Ablikim,Hai-Bo Li,Liaoyuan Dong}, GAMS-group at CERN and BNLE852-collaboration \cite{G.S. Adams,E.I. Ivanov,J. Kuhn,Daniel S. Carman} had observed  few states (below 2.5 GeV), may have exotic structure and quantum numbers, required theoretical attentions. In such aspect, the spectroscopy of mesons are playing a dominant role in the study of the QCD, especially, the light meson sector which carries the information of nonperturbative regime. The scalar mesons in the light mass regime may give the clue for the symmetry breaking, as they have the same quantum numbers as of vacuum. The difficulties with the study of light mesons sector are their large decay widths as well as their appearance does not come with narrow isolated peaks.   \\

\begin{table}[]
\begin{center}
\caption{Experimentally observed states (in MeV) \cite{21}}
\label{tab-3}
\begin{tabular}{ c c c c c}
\hline 
States &Mass& Possible & Compared \\
 & (MeV))& $I^G (J^{PC})$ &  dimesonic  \\
  & &  & states  \\
  \hline 
$f_0(980)$&990$\pm$20 &$0^+(0^{++})$ &$K-\bar{K}$   \\
$h_1(1380)$&1386$\pm$19  &$0^-(1^{+-})$ &$K-\bar{K^*}$ \\
$a_0(1450)$&1474$\pm$19 &$1^-(0^{++})$ &$\rho-\bar{\omega}$ \\
$f_0(1500)$&1505$\pm$06 &$0^+(0^{++})$ &$\rho-\bar{\rho}$ \\
$f'_2(1525)$&1525$\pm$05  &$0^+(2^{++})$ &$\rho-\bar{\rho}$ \\
$f_2(1565)$&1562$\pm$13  &$0^+(2^{++})$ &$\omega-\bar{\omega}$ \\
$a_2(1700)$& $1722\pm16$  &$1^-(2^{++})$ &$K^*-\bar{K^*}$ \\
$f_0(1710)$& $1722^{+06}_{-05}$  &$0^+(0^{++})$ &$K^*-\bar{K^*}$ \\
$f_2(1810)$& 1815$\pm$12 &$0^+(2^{++})$ &$\phi-\bar{\omega}$ \\
$f_2(2010)$&$2011_{-80}^{+60}$  &$0^+(2^{++})$ &$\phi-\bar{\phi}$ \\
\hline 
\end{tabular}
\end{center}
\end{table}
  
In this paper, we have studied the molecule like structure of mesons in the light flavour (u,d,s) sector. We have listed many possible candidates for the exotic states in Table-1 (for reference). The multiquark states have been studied and predicted previously in Bag model and in nonrelativistic potential models \cite{5,Weinstein,6,Weinstein2,7,11,12,14,15,16}.  The molecular like structures have been studied in various theoretical approaches like potential models \cite{11,12,14,15,16,17,22,41,Gui-Jun Ding}, gauge invariant model \cite{Tanja Branz}, QCD sum rules  \cite{Jian-Rong Zhang,Jian-Rong Zhang2}, Bethe-Salpeter equation approach \cite{26} and field theory approach \cite{C. Hidalgo-Duque}. Recentaly, the molecular-like states are predicted in the framework of chiral SU(3) quark model \cite{W L Wang}, Ref.  \cite{Ju-Jun Xie} have studied  the photoproduction of the tensor states. \\

We have studied the molecular dimesonic (meson-antimeson) systems in the potential model framework.  The multiquark structure (four quark state) could be formed by two ways, (i) no subcomponent of system could make color neutral, probably known as tetraquark. (ii) the quark-antiquark subcomponent (meson) is being color neutral and same for the other subcomponent (antimeson). In the second case the both color neutral subcomponent makes bound state, say dimesonic state.  The dimesonic system consists the meson-antimeson bound state, just like deuteron (bound state of the proton and neutron). By taking this approximation, we have employed various interaction potentials like Yukawa like potential, one pion exchange and sigma exchange potential. The meson-antimeson system have been studied previously with the same approximation by T$\ddot{o}$rnqvist \cite{22,41} and introduced it as a $'duesons'$. The mass of the dimesonic state should be less than the sum of the mass of the two respective mesons and the binding energy of the system should be small. The theoretical calculation of this binding energy of dimesonic states is very complex. As such, it deals with the delicate cancellation of short range and long range repulsion and attraction respectively. \\

Recently, We have successfully used the Hellmann (Coulomb$+$Yukawa) potential with one pion exchange potential to study the molecular binding of meson-antimeson systems \cite{Rathaud}. In the present study, we replaced these pseudopotential by Yukawa like potential with one pion exchange and sigma exchange potential. In contrast with \cite{Rathaud}, to incorporate the meson exchange theory in a more consistent way, we have used the sigma exchange with Yukawa- like potential. The Yukawa like potential creates screening effects through a parameter appeared in the potential somehow control the coulombic interaction. For instance, we assume it as for gulonic field and the screening parameter as a color screening parameter. In such a way, our model carried only one fitting parameter.\\

With this potential, in the Ritz variational scheme and by using the hydrogenlike trial wave function, we have solved the Schroedinger equation and extract the ground state energy eigenvalue. The aim of the study is to obtained the mass spectra of dimesonic systems in the one boson exchange ($\pi,\sigma$) model with the Yukawa like potential. To test the interaction potential mechanism used by us, we apply it first to  the state $f_{0}(980)$. The  state $f_{0}(980)$ has widely been believed to as a $K-\overline{K}$ molecule in the literature. We calculate the mass spectra and binding energy  for the  $f_{0}(980)$.  
Due to the parity violation, the one pion exchange potential could not be applied to the dimesonic systems consist of two pseudoscalars. As such, these systems interacting through sigma exchange with the Yukawa like potential. The whole model carry only one fitting parameter appeared in the Yukawa like potential. We fit this parameter to get the approximate binding energy of the  state $f_{0}(980)$ and according to this binding energy the digamma width have been found very close to the experimental results \cite{21}.  Thus, we have fixed this parameter for all other dimeosnic combinations for their mass spectra.  
We have calculated the digamma decay width by using the formula given by Ref.\cite{18}. 
In the dimesonic states, we have used meson and anti-meson constituents which are colour singlets, so confinement potential is not required as required in quark anti-quark system \cite{Ruizhang}. In this molecular system the kinetic energy of the constituent mesons in the bound state is also small, so that the relativistic effects can be neglected and the present method represents the same \cite{Ruizhang}. By considering the Pseudoscalar and Vector mesons for dimesonic states, we have three combinations (i) Pseudoscalar-Pseudoscalar (PP-states) (ii) Pseudoscalar-Vector (PV-states) (iii) Vector-Vector (VV-states).   \\

The paper is arranged as follows, after the introduction, we present the framework and model of the present study, in the subsection we have discussed the digamma widths of the systems. In the third section, we have discussed the results and the last section dedicated for conclusions.\\

\section{Theoretical framework}
\label{sec:1}
The variational scheme is employed to solve the Schrödinger equation and to get the value of  variational parameter for each state, the virial theorem is used.
The Hamiltonian of the dimesonic (meson-antimeson) system is given by \cite{9,9a,8,8a}
\begin{equation}
H=\sqrt{P^2+m_{h1}^{2}}+\sqrt{P^2+m_{h2}^2}+V(r)
\end{equation}
where $m_{h1}$ and $m_{h2}$ are masses of mesons, P is the relative momentum of two mesons and V(r) is the molecular interaction potential of the dimesonic system. In the present study, we have dealt with the light flavour mesons (except pions). To incorporate the relativistic correction to the kinetic energy, we have expanded the kinetic energy term of the Hamiltonian up to $\cal{O}$($P^{8}$)
\begin{eqnarray}
K.E. &=& \frac{P^{2}}{2}\left(\frac{1}{m_{h1}} +{ \frac{1}{m_{h2}}}\right) - \frac{P^{4}}{8}  \left(\frac{1}{m_{h1}^{3}} +{ \frac{1}{m_{h2}^{3}}}\right) \nonumber \\ & &
 + \frac{P^{6}}{16}  \left(\frac{1}{m_{h1}^{5}} +{ \frac{1}{m_{h2}^{5}}}\right)  + {\cal{O}}(P^{8})
\end{eqnarray}
The interaction potential of meson-antimeson system is consisted the Yukawa like potential, one pion exchange potential (OPEP) ($V_{\pi}$) and sigma exchange potential ($V{}_{\sigma}$). Whereas, the spin dependent potential $V{}_{SD}$ added perturbatively. 

\begin{equation}
V(r)= V(r_{12})+V{}_{\sigma}+V{}_{\pi}
\end{equation}
\begin{equation}
V(r_{12})=-\frac{k{}_{mol}}{r_{12}}e{}^{-\frac{C^{2}{r_{12}}^{2}}{2}}
\end{equation}
$ r_{12}$ is the relative co-ordinates of meson-antimeson system, $k{}_{mol}$ is the residual strength of strong running coupling constant and C is the effective color screening parameter of the confined gluon \cite{11,12}. The dimesonic molecular interaction is consist  with the complicated  interaction of short range repulsion and long range attraction. The Yukawa like potential and sigma exchange potential cares for short and mid range interaction while one pion exchange potential  take care for long range part. 
The value of the $k{}_{mol}$  (running coupling constant) is determined through,
\begin{equation}
k{}_{mol} = \frac{4\pi}{(11-\frac{2}{3}n_{f})ln\frac{M^{2}+ {M_{B}}^{2}}{\Lambda^{2}}}
\end{equation}
where M = 2$m_{h1}$ $m_{h2}$/($m_{h1}$+$m_{h2}$), $M_{B}$=0.95 GeV and $\Lambda$=0.413 GeV \cite{Ebert2009} whereas $ n_{f}$ is number of flavours. One pion exchange potential (OPEP) and $\sigma$ exchange potential \cite{19,Furuichi,20} is used, based on the assumption that molecular like structure of dimesonic system is being deuterium like structure of nucleon \cite{17}. The conventional OPE potential is consist with the tensor interaction term which plays a dominant role in NN scattering \cite{VijayR,VijayR1}. If two hadrons are in L=0 (ground state) state, then the matrix element of tensor operator becomes zero \cite{23}. The sigma exchange includes spin-orbital interactions and as we dealt with the system of S-wave ($L=0$), the term with spin-orbital coupling vanishes. The nonrelativistic form of the pion and sigma exchange potential is used here  \cite{19,Furuichi,20}.  
\begin{center}
\begin{eqnarray}
V{}_{\pi} & = & \frac{1}{3}\frac{g^{2}{}_{8}}{4\pi}\left(\frac{m{}_{\pi}^{2}}{4m_{1}m_{2}}\right)\left(\tau{}_{1}\cdot\tau{}_{2}\right)
\left(\sigma{}_{1}\cdot\sigma{}_{2}\right) \nonumber \\ & & \left(\frac{e{}^{-m_{\pi}r_{12}}}{r_{12}}-\left(\frac{\Lambda{}_{\pi}}{m_{\pi}}\right)^{2}\frac{e{}^{-\Lambda_{\pi}r_{12}}}{r_{12}}\right)
\end{eqnarray}

\par\end{center}
\begin{center}
\begin{equation}
V{}_{\sigma}=\frac{g^{2}{}_{8}}{4\pi} \left(\frac{e{}^{-m_{\sigma}r_{12}}}{r_{12}}-\frac{e{}^{-\Lambda_{\sigma}r_{12}}}{r_{12}}\right)
\end{equation}
\par\end{center}

where  $g$=$g_{8}$ is a quark-meson coupling constant for $\pi, K, $ and $ \eta$. The coupling constant for sigma  meson is taken to be the same as for pseudoscalar meson, $g{}_{8}$=0.69, is taken from \cite{19,Furuichi,20}. The {$\bf\tau$} and {$\bf\sigma$} are isospin, spin matrices respectively. $\Lambda{}_{M}$= $km{}_{M}$ ($m_{M}$-exchange meson mass) is the form factor, appears due to the dressing of quarks and assumed to be depends on the exchange boson masses. Where $m_{i}$ and $m{j}$ are the constituent masses in  GeV, $ m_{\pi}$=0.1349 GeV, $m_{\sigma}$=0.675 GeV and  $k$=2.2. For PV states the values of spin-isospin factors have been taken as $\left(\tau{}_{i}\cdot\tau{}_{j}\right)\left(\sigma{}_{i}\cdot\sigma{}_{j}\right)\ $=-3,1 for I=0,1. While for VV states  values taken as $\left(\tau{}_{i}\cdot\tau{}_{j}\right)\left(\sigma{}_{i}\cdot\sigma{}_{j}\right)\ $=-6,-3,3 for isospin I=0 and spin S=0,1,2 respectively. Whereas for isospin I=1 and spin S=0,1,2 it takes values $\left(\tau{}_{i}\cdot\tau{}_{j}\right)\left(\sigma{}_{i}\cdot\sigma{}_{j}\right)\ $= 2,1,-1. The values of spin-isospin factor are from ref.\cite{22,41} and the other parameters used in pion exchange potential are from Ref. \cite{19,Furuichi,20,23}.\\

\begin{table}[t]
\begin{center}
\caption{Masses of mesons (in MeV) \cite{21}}
\label{tab-3}
\begin{tabular}{ c c c c c c c c c}
\hline 
Meson  &  $K$ & $\eta$ & $\eta'$ & $\rho$ & $\omega$ &$K{}^{*}$ &  $\phi$ \tabularnewline
Mass &  497.6 & 547.8 & 957.7 & 775.4 & 782.6  & 895.9& 1019.4  \tabularnewline
\hline 
\end{tabular}
\end{center}
\end{table}

The spin dependent interaction potential added separately \cite{11,12}
\begin{equation}
V_{SD}= \frac{8}{9}\frac{K_{mol}}{m_{h1}m_{h2}}{S_{1}\cdot{S_{2}}}{\vert\psi(0)\vert}^{2}
\end{equation}
Where  the spin factor ${\bf S_{1}\cdot S_{2}}$ can be found by general formula ${\bf S_{1}\cdot S_{2}}=\frac{1}{2}[({\bf S_{1}}+{\bf S_{2}})^{2}-{\bf S_{1}}^{2}-{\bf S_{2}}^{2}]$ \cite{11}. We have used the hydroginic like trial wave function such as,
\begin{eqnarray}
R_{nl}(r)&=&\left(\frac{\mu{}^{3}(n-l-1)!}{2n(n+l)^{3}!}\right)^{\frac{1}{2}}\left(\mu r\right)^{l}\nonumber \\ & & e{}^{\frac{-\mu r}{2}}L_{n-l-1}^{2l+1}(\mu r)
\end{eqnarray}

\begin{table*}[t]
\begin{center}
\caption{Spin-spin and Iso-spin and $J^{PC}$ for some states}
\begin{tabular}{ c c c c c c c c c c c c c c }
\hline 
$L{}_{1}$ & $L{}_{2}$ & $S{}_{1}$ & $S{}_{2}$ & $L{}_{12}$ & $S{}_{12}$ & J & $I{}_{1}$ & $I{}_{2}$ & $I$ & $J{}^{PC}$ & $I{}^{G}$ & $^{2S+1}X_{J}$ & States\tabularnewline
\hline 

0 & 0 & 0 & 0 & 0 & 0 & 0 & 0 & 0 & 0 & $0{}^{++}$ & $0{}^{+}$ & $^{1}S_{0}$ & $f{}_{0}(980)$\tabularnewline

0 & 0 & 1 & 0 & 0 & 1 & 1 & 0 & 0 & 0 & $1{}^{+-}$ & $0{}^{-}$ & $^{3}S_{1}$ & $h{}_{1}(1380)$\tabularnewline

0 & 0 & 0 & 0 & 0 & 0 & 0 & 0 & 1 & 1 & $0{}^{++}$ & $1{}^{-}$ & $^{1}S_{0}$ & $f{}_{0}(1500)$\tabularnewline

0 & 0 & 1 & 0 & 0 & 1 & 1 & 1 & 0 & 1 & $1{}^{+-}$ & $1{}^{+}$ & $^{3}S_{1}$ & \tabularnewline

0 & 0 & 1 & 1 & 0 & 0 & 0 & 0 & 0 & 0 & $0{}^{++}$ & $0{}^{+}$ & $^{1}S_{0}$ & $f{}_{0}(1710)$\tabularnewline

 &  &  &  &  & 1 & 1 &  &  &  & $1{}^{+-}$ & $0{}^{-}$ & $^{3}S_{1}$ & \tabularnewline

 &  &  &  &  & 2 & 2 &  &  &  & $2{}^{++}$ & $0{}^{+}$ & $^{5}S_{2}$ & $f{}_{2}(1810)$\tabularnewline
\hline 
\end{tabular}
\end{center}
\end{table*}

where $\mu$ is the variational parameter and $ L_{n-l-1}^{2l+1}(\mu r)$ is the Laguerre polynomial. We have fixed the color screening parameter C = 0.15 GeV for all the dimesonic combinations. The experimental (PDG) masses of the mesons are used for the present study, tabulated in Table-2. We have obtained the expectation value of Hamiltonian as 
\begin{equation}
H \psi = E \psi
\end{equation}
by employing the Ritz variational scheme for mass calculation and
$\mu$ is determined for each state using the virial theorem \cite{9,8}.
\begin{equation}
\left <KE.\right >  =  \frac{1}{2} \left< \frac{rdV(r)}{dr}\right>
\end{equation}
The one pion exchange potential for the meson-antimeson system in Eq.(6) is spin-isospin dependent. The spin-isospin dependency of one pion exchange have been  discussed by T$\ddot{o}$rnqvist for $'deusons'$ \cite{22,41}. We get attractive channel for (S,I) = (1,0) for PV-states while repulsive for (S,I) = (1,1) channel. In VV-states, we got an attractive channel for (S,I) = (0,0)(1,0)(2,0)  while repulsive for (S,I) = (0,1)(1,1)(2,0). The two Pseudoscalar could not be bound by Pseudoscalar due to parity violation \cite{17,22,41}, for such PP-states, we have not considered the one pion exchange potential. For the PV- states, the  pion as a constituent is too light and gives large kinetic energy which is difficult to beat by potential to bind a molecule \cite{41}. Thus, we have not considered the pion for dimesonic system, as it required relativistic treatment. The calculated masses are close to experimental measurements(PDG), tabulated in Table-4.\\ 

Since angular momentum, spin, isospin, parity all are conserved in strong interaction, be a good quantum number for  dimesonic system. Thus, we can apply a usual coupling rule to dimesonic states. The orbital angular momentum of meson and antimeson are $L_{1}$ and $L_{2}$ respectively. In same way spin denotes by $S_{1}$, $S_{2}$ and isospin $I_{1}$, $I_{2}$. Employing the coupling rules, one has the relative orbital momentum and total spin of system $L_{12}$ and $S_{12}$. Whereas, the total angular momentum to be {\bf{J}} = { $\bf L_{12}$}+{  $\bf S_{12}$}. In the similar way,  we have {\bf I} = {$\bf I_1$}+{ $\bf I_2$}. The parity of the two particle system (meson - antimeson) is given by P = $P_{1}$ $P_{2}$ $(-1)^{L_{12}}$ \cite{15,16} while the charge conjugation be C = $(-1)^{L_{12}+S{12}}$ \cite{40} and G-parity define as G = $(-1)^{L_{12}+S_{12}+I}$. While one can refer  ref. \cite{15,16,40} for more detailed discussion about quantum numbers for normal mesons as well as for exotic states. The spin, isospin, angular momentum for some states  are listed in Table-3.

\begin{table*}[]

\begin{center}
\caption {Binding energy and Mass spectra of dimesonic  ($q\overline{q}-\overline{q}q$)  systems  with their $J^{PC}$ values.}
\label{tab-1}

\begin{tabular}{ccccccc}
\hline 
System & $I{}^{G}(J^{PC})$ & $\mu$ & B.E.& Mass & Expt. \cite{21} & State\tabularnewline
 
 &  & (GeV)  & (MeV) & (GeV) & (GeV)  & \tabularnewline
\hline
PP-states &&&&&&\\

$K-\overline{K}$  & $0{}^{+}$ $(0{}^{++})$ & 0.2823  & -09.79  & 0.985 & $0.990\pm0.020$ & $f{}_{0}(980)$\tabularnewline

$\eta-\overline{\eta}$ & $0{}^{+}$ $(0{}^{++})$ & 0.3065  & -12.57 & 1.083 & & \tabularnewline

$\eta-\overline{\eta}'$ & $0{}^{+}$ $(0{}^{++})$ & 0.3623  & -19.01 & 1.486 &  &   \tabularnewline
 
$\eta'-\overline{\eta}'$& $0{}^{+}$ $(0{}^{++})$  & 0.4239  & -24.84  & 1.890 &  &   \tabularnewline
\hline
PV-states &&&&&&\\
$\eta-\overline{\rho}$ & $1{}^{+}$ $(1{}^{+-})$ & 0.3410 & -16.57 & 1.306 &  &    \tabularnewline

$\eta-\overline{\omega}$ & $0{}^{-}$ $(1{}^{+-})$ & 0.3539  & -18.20 & 1.312 & &   \tabularnewline

$K-\overline{K}^{*}$ &$0{}^{-}$ $(1{}^{+-})$ & 0.3509 & -17.84  & 1.371 & $1.386\pm0.019$ &  $h_{1}(1380)$  \tabularnewline

$K-\overline{K}^{*}$ & $1{}^{+}$ $(1{}^{+-})$  & 0.3394  & -16.38 & 1.373 &  &   \tabularnewline

$\eta-\overline{\phi}$ & $0{}^{-}$ $(1{}^{+-})$ & 0.4070  & -26.90  & 1.540 &  &  \tabularnewline

$\eta'-\overline{\rho}$ & $1{}^{+}$ $(1{}^{+-})$ & 0.4016  & -22.91  & 1.710 &  &   \tabularnewline
 
$\eta'-\overline{\omega}$ & $0{}^{-}$ $(1{}^{+-})$ & 0.4117  & -24.20 & 1.716 &  &   \tabularnewline
 
$\eta'-\overline{\phi}$& $0{}^{-}$ $(1{}^{+-})$ & 0.4712  & -32.96  & 1.944 & &   \tabularnewline

\hline 
VV-states &&&&&&\\
$\rho-\overline{\rho}$ & $0{}^{+}$ $(0{}^{++})$ & 0.4000  & -23.44 & 1.472 & $1.505\pm0.006$ &  $f{}_{0}(1500)$\tabularnewline

$\rho-\overline{\rho}$ & $0{}^{-}$ $(1{}^{+-})$ & 0.3921  & -22.37 & 1.502 &  &  \tabularnewline
 
$\rho-\overline{\rho}$ & $0{}^{+}$ $(2{}^{++})$ & 0.3768  & -20.36 & 1.553 & $1.525\pm0.005$ &  $f{}_{2}'(1525)$ \tabularnewline
 
$\rho-\overline{\rho}$ & $1{}^{-}$ $(0{}^{++})$ & 0.3793  & -20.68 & 1.483 &  & \tabularnewline
 
$\rho-\overline{\rho}$ & $1{}^{+}$ $(1{}^{+-})$ & 0.3818  & -21.01 & 1.506 &  &  \tabularnewline

$\rho-\overline{\rho}$ & $1^{-}$ $(2{}^{++})$ & 0.3869 &  -21.68 & 1.554 &  &   \tabularnewline
\hline 
$\omega-\overline{\omega}$ & $0{}^{+}$ $(0{}^{++})$ & 0.4017  & -23.60 & 1.487 &  &  \tabularnewline

$\omega-\overline{\omega}$ & $0{}^{-}$ $(1{}^{+-})$ & 0.3939  & -22.55 & 1.517 &  & \tabularnewline
 
$\omega-\overline{\omega}$ & $0{}^{+}$ $(2{}^{++})$ & 0.3787 & -20.55 & 1.567 & $1.562\pm0.013$ & $f{}_{2}(1565)$  \tabularnewline
\hline 
$K^{*}-\overline{K}^{*}$ & $0{}^{+}$ $(0{}^{++})$ & 0.4256  & -25.70 & 1.719 &  $1.722_{-5}^{+6}$ &  $f{}_{0}(1710)$ \tabularnewline
 
$K^{*}-\overline{K}^{*}$ & $0{}^{-}$ $(1{}^{+-})$ & 0.4188  & -24.79 & 1.744 &  &   \tabularnewline

$K^{*}-\overline{K}^{*}$ & $0{}^{+}$ $(2{}^{++})$ & 0.4055  & -23.06 & 1.789 &  &   \tabularnewline
 
$K^{*}-\overline{K}^{*}$ & $1{}^{-}$ $(0{}^{++})$ & 0.4077  & -23.34 & 1.727 &  &  \tabularnewline
 
$K^{*}-\overline{K}^{*}$ & $1{}^{+}$ $(1{}^{+-})$ & 0.4099  & -23.63 & 1.747 &  &   \tabularnewline

$K^{*}-\overline{K}^{*}$ & $1^{-}$ $(2{}^{++})$ & 0.4143  & -24.20 & 1.789 & $1.722\pm0.016 $ & $a{}_{2}(1700)$  \tabularnewline
\hline 

$\phi-\overline{\phi}$ & $0{}^{+}$ $(0{}^{++})$ & 0.4462  & -27.08 & 1.972 &  &   \tabularnewline

$\phi-\overline{\phi}$ & $0{}^{-}$ $(1{}^{+-})$ & 0.4402  & -26.31 & 1.993 &  &   \tabularnewline
 
$\phi-\overline{\phi}$ & $0{}^{+}$ $(2{}^{++})$ & 0.4285  & -24.82 & 2.031 & $2.011^{+0.06}_{-0.08}$ &  $f{}_{2}(2010)$ \tabularnewline
\hline 
$\rho-\overline{\omega}$ & $1{}^{-}$ $(0{}^{++})$ & 0.3803  & -20.78 & 1.490 & $ 1.474\pm0.019$ & $a{}_{0}(1450)$ \tabularnewline

$\rho-\overline{\omega}$ & $1{}^{+}$ $(1{}^{+-})$ & 0.3828  & -21.10 & 1.513 &  &   \tabularnewline
 
$\rho-\overline{\omega}$ & $1^{-}$ $(2{}^{++})$ & 0.3879  & -21.77 & 1.561 & &   \tabularnewline
 \hline
$\phi-\overline{\rho}$ & $1{}^{-}$ $(0{}^{++})$ & 0.4388  & -29.67 & 1.708 &  &   \tabularnewline
 
$\phi-\overline{\rho}$ & $1{}^{+}$ $(1{}^{+-})$ & 0.4411  & -30.00 & 1.736 &  & \tabularnewline
 
$\phi-\overline{\rho}$ & $1^{-}$ $(2{}^{++})$ & 0.4457  & -30.68 & 1.793 &  &  \tabularnewline
 \hline
$\phi-\overline{\omega}$ & $0{}^{+}$ $(0{}^{++})$ & 0.4584 & -32.53 & 1.705 & &   \tabularnewline
 
$\phi-\overline{\omega}$ & $0{}^{-}$ $(1{}^{+-})$ & 0.4514 &  -31.46 & 1.740 &  &  \tabularnewline
 
$\phi-\overline{\omega}$ & $0{}^{+}$ $(2{}^{++})$ & 0.4376 &  -29.43 & 1.800 & $1.815\pm0.012 $ & $ f_{2}(1810)$ \tabularnewline
\hline 

\end{tabular}
\end{center}

\end{table*}

\begin{table*}[]
\begin{center}
\caption {Root mean square radius and  digamma decay width of dimesonic  ($q\overline{q}-\overline{q}q$) systems.}
\label{tab-1}
\begin{tabular}{ccccccc}
\hline 
System & $I{}^{G}(J^{PC})$ &  R(0) & $\sqrt{<r^{2}>}$ & $\Gamma_{\gamma\gamma}$ & Exp. \cite{21} & State\tabularnewline
 
 &  & $(GeV{}^{\frac{3}{2}})$  & (fm) & KeV & KeV & \tabularnewline
\hline
PP-states &&&&&&\\

$K-\overline{K}$  & $0{}^{+}$ $(0{}^{++})$ & 0.1060 & 02.42 & 0.2260 & $0.29_{-0.06}^{+0.07}$ & $f{}_{0}(980)$\tabularnewline

$\eta-\overline{\eta}$ & $0{}^{+}$ $(0{}^{++})$ & 0.1200 & 02.23 & 0.2882 & & \tabularnewline

$\eta-\overline{\eta}'$ & $0{}^{+}$ $(0{}^{++})$  & 0.1542 & 01.88 & 0.2556 & &  \tabularnewline
 
$\eta'-\overline{\eta}'$& $0{}^{+}$ $(0{}^{++})$  & 0.1951 & 01.61 & 0.3274 &  &  \tabularnewline
\hline
PV-states &&&&&&\\
$\eta-\overline{\rho}$ & $1{}^{+}$ $(1{}^{+-})$  & 0.1408 & 02.00 & 0.5082 & &   \tabularnewline

$\eta-\overline{\omega}$ & $0{}^{-}$ $(1{}^{+-})$ & 0.1489 & 01.93 & 0.5340 & &  \tabularnewline

$K-\overline{K}^{*}$ &$0{}^{-}$ $(1{}^{+-})$  & 0.1469 & 01.94 & 0.1676 & & $h_{1}(1380)$  \tabularnewline

$K-\overline{K}^{*}$ & $1{}^{+}$ $(1{}^{+-})$ & 0.1398 & 02.01 & 0.1548 & &  \tabularnewline

$\eta-\overline{\phi}$ & $0{}^{-}$ $(1{}^{+-})$ & 0.1836 & 01.67 & 0.2480 & &  \tabularnewline

$\eta'-\overline{\rho}$ & $1{}^{+}$ $(1{}^{+-})$ & 0.1800 & 01.70 & 1.5308 & &  \tabularnewline
 
$\eta'-\overline{\omega}$ & $0{}^{-}$ $(1{}^{+-})$ & 0.1868 & 01.65 & 1.5159 & &  \tabularnewline
 
$\eta'-\overline{\phi}$& $0{}^{-}$ $(1{}^{+-})$ & 0.2287 & 01.45 & 1.5617 & &  \tabularnewline

\hline 
VV-states &&&&&&\\
$\rho-\overline{\rho}$ & $0{}^{+}$ $(0{}^{++})$  & 0.1789 & 01.70 & 0.4233 & & $f{}_{0}(1500)$\tabularnewline

$\rho-\overline{\rho}$ & $0{}^{-}$ $(1{}^{+-})$  & 0.1736 & 01.74 & 0.5396 & & \tabularnewline
 
$\rho-\overline{\rho}$ & $0{}^{+}$ $(2{}^{++})$  & 0.1635 & 01.81 & 0.7067 & $0.081\pm0.009$ & $f{}_{2}'(1525)$ \tabularnewline
 
$\rho-\overline{\rho}$ & $1{}^{-}$ $(0{}^{++})$  & 0.1652 & 	01.80 & 0.4344 & & \tabularnewline
 
$\rho-\overline{\rho}$ & $1{}^{+}$ $(1{}^{+-})$  & 0.1668 & 01.79  & 0.5413 & & \tabularnewline

$\rho-\overline{\rho}$ & $1^{-}$ $(2{}^{++})$ & 0.1702 & 01.76 & 0.7238 & &  \tabularnewline
\hline 
$\omega-\overline{\omega}$ & $0{}^{+}$ $(0{}^{++})$ & 0.1800 & 01.70 & 0.2120 & & \tabularnewline

$\omega-\overline{\omega}$ & $0{}^{-}$ $(1{}^{+-})$ & 0.1748 & 01.73 & 0.2739 & & \tabularnewline
 
$\omega-\overline{\omega}$ & $0{}^{+}$ $(2{}^{++})$ & 0.1648 & 01.80 & 0.7167 & $0.70\pm0.14 $ & $f{}_{2}(1565)$  \tabularnewline
\hline 
$K^{*}-\overline{K}^{*}$ & $0{}^{+}$ $(0{}^{++})$ & 0.1963 & 01.60 & 0.2690 &  & $f{}_{0}(1710)$ \tabularnewline
 
$K^{*}-\overline{K}^{*}$ & $0{}^{-}$ $(1{}^{+-})$ & 0.1916 & 01.63 & 0.3320 & &  \tabularnewline

$K^{*}-\overline{K}^{*}$ & $0{}^{+}$ $(2{}^{++})$ & 0.1826 & 01.68 & 0.6618 & &  \tabularnewline
 
$K^{*}-\overline{K}^{*}$ & $1{}^{-}$ $(0{}^{++})$ & 0.1840 & 01.67 & 0.2740 & &  \tabularnewline
 
$K^{*}-\overline{K}^{*}$ & $1{}^{+}$ $(1{}^{+-})$ & 0.1855 & 01.66 & 0.3326 & &  \tabularnewline

$K^{*}-\overline{K}^{*}$ & $1^{-}$ $(2{}^{++})$ & 0.1885 & 01.64 & 0.6843 &$0.30\pm0.05$ &  $a{}_{2}(1700)$ \tabularnewline
\hline 

$\phi-\overline{\phi}$ & $0{}^{+}$ $(0{}^{++})$ & 0.2107 & 01.53 & 0.3319 & &  \tabularnewline

$\phi-\overline{\phi}$ & $0{}^{-}$ $(1{}^{+-})$ & 0.2065 & 01.55 & 0.3939 & &  \tabularnewline
 
$\phi-\overline{\phi}$ & $0{}^{+}$ $(2{}^{++})$ & 0.1984 & 01.59 & 0.6486 & &  $f{}_{2}(2010)$ \tabularnewline
\hline 
$\rho-\overline{\omega}$ & $1{}^{-}$ $(0{}^{++})$ & 0.1658 & 01.79 & 0.2170 & &  $a{}_{0}(1450)$ \tabularnewline

$\rho-\overline{\omega}$ & $1{}^{+}$ $(1{}^{+-})$ & 0.1674 & 01.78 & 0.2734 & &  \tabularnewline
 
$\rho-\overline{\omega}$ & $1^{-}$ $(2{}^{++})$ & 0.1708 & 01.76 & 0.7287 & &  \tabularnewline
 \hline
$\phi-\overline{\rho}$ & $1{}^{-}$ $(0{}^{++})$ &.2056 & 01.55 & 0.2942 & &  \tabularnewline
 
$\phi-\overline{\rho}$ & $1{}^{+}$ $(1{}^{+-})$ & 0.2072 & 01.54 & 0.3825 & & \tabularnewline
 
$\phi-\overline{\rho}$ & $1^{-}$ $(2{}^{++})$ & 0.2104 & 01.53 & 0.8025  & & \tabularnewline
 \hline
$\phi-\overline{\omega}$ & $0{}^{+}$ $(0{}^{++})$ & 0.2194 & 01.49 & 0.2835 & &  \tabularnewline
 
$\phi-\overline{\omega}$ & $0{}^{-}$ $(1{}^{+-})$ & 0.2144 & 01.51 & 0.3767 & & \tabularnewline
 
$\phi-\overline{\omega}$ & $0{}^{+}$ $(2{}^{++})$ & 0.2047 & 01.56 & 0.8043 & & $ f_{2}(1810)$ \tabularnewline
\hline 
\end{tabular}
\end{center}
\end{table*}

\subsection{Digamma width}
\label{sec:3}
The digamma decay width have been calculated on hadron loops employing an expansion in the range for force while the leading term assumes to a point like vertex. The formula with leading order correction \cite{18} is being
\begin{eqnarray}
&\Gamma_{\gamma\gamma}&=\xi^{2}\left(\frac{\alpha}{\pi}\right)^{2}\sqrt{m\varepsilon}\left(\frac{2m}{m_{sa}}\right) \nonumber \\ & & \left[\left(\frac{2m}{m_{sa}}\right)^{2} 
arcsin^{2} \left(\frac{m_{sa}}{2m}\right)-1\right]^{2}
\end{eqnarray}

\begin{equation}
\Gamma_{\gamma\gamma}(\beta)= \Gamma_{\gamma\gamma} \left(1+ \xi \frac{4m_{h1}m_{h2}-m_{sa}^{2}}{\beta^{2}}\right)
\end{equation}
$\Gamma_{\gamma\gamma} $ is given by eq.(12)	
\begin{equation}
m_{sa}=2m-\varepsilon
\end{equation}
where $\alpha=e^{2}/4\pi$ indicate fine-structure constant and the
factor $\xi$ =$\frac{1}{\sqrt{2}}$, the $\beta = 0.8 $ GeV, m is the mass of constituent meson, $m_{sa}$ is the mass of molecular meson and $ \varepsilon$ is binding energy. The digamma decay widths are tabulated in Table-5. 

\section{Results and Discussion}
\label{sec:3}
We have employed the semirelativistic approach to calculate the masses and digamma decay width of the proposed dimesonic molecular states. Here, we have considered the three combination PP-states, PV-states and VV-states. Our results are tabulated in Table-4 $\&$5.
The Yukawa-like potential with sigma exchange and one pion exchange potential are used as an interaction potential for dimesonic systems. We have found that the contribution from Yukawa-like potential being approximately 80$\sim$85$\%$ (attractive), sigma exchange 12$\sim$15$\%$ (repulsive) and one pion exchange uo to $3\%$ (attractive or repulsive depending upon spin-isospin channel) to the net potential strength.\\

We have applied our model to $f{}_{0}(980)$  which is widely believed to be a $K-\overline{K}$ molecule in literatures \cite{21}. The color screening parameter C is the only the parameter which has fitted once, to get the approximate mass of the state $f{}_{0}(980)$. The calculated mass is found $985$ MeV with approximately 9.8 MeV binding energy and both are close to experimental measurements \cite{21}. In the Ref. \cite{Rathaud}, the binding energy was found nearly $20$ MeV which is almost twice then in the present study (see Table-3 Ref.\cite{Rathaud}), same with other dimesonic states. The digamma width and mass is being found close in agreement with experimental results \cite{21}. Thus, we have used this model to study other dimesonic combinations. The various dimesonic states are very close to the experimentally observed states in accordance to their masses. \\

The molecular state should have small binding and large radius compared to the size of their constituents. We have found radius of all combinations near $\sim 2$ fm to $\sim 4$ fm with small binding energy, see Table-4 (negative sign shows the bound state). The Various states like $h_{1}(1380)$ in PV-state,  $f_{0}(1500)$, $f_{2}'(1525)$, $f_{2}(1565)$, $f_{0}(1710)$, $a_{2}(1700)$, $a_{0}(1450)$, $f_{2}(1810)$, $f_{2}(2010)$ in VV-states are compared with dimesonic states.  The masses of the states $f{}_{0}(980)$, $h_{1}(1380)$, $f_{2}(1565)$, $f_{0}(1710)$, $a_{0}(1450)$, $f_{2}(2010)$, $f_{2}(1810)$ are in excellent agreement with the PDG listing \cite{21}. Moreover, some states like $f_{0}(1500)$ and $a_{2}(1700)$ compared with $\rho-\overline{\rho}$ and $K^{*}-\overline{K^{*}}$ having 20 to 50 MeV deference in mass compared to experimental measurement \cite{21}. The quantum numbers for the dimesonic system are assigned as discussed in the previous section. The digamma decay width for the states $f{}_{0}(980)$ and  $f{}_{2}(1565)$ are in agreement with experimental values, while for the state  $a_{2}(1700)$  has been overestimated,
 see Table-5.\\
 
\section{Conclusions}
On the basis of mass spectra and quantum numbers, we are able to identify states having promising non $ q\overline{q}$  structures. The calculated masses and digamma widths of the compared states are close to the experimental measurements \cite{21} except few states like $f_{0}(1500)$, $f_{2}'(1525)$ and $a_{2}(1700)$ (see Table-3 $\&$4). Moreover, the states $f_{0}(1500)$, $f_{2}'(1525)$ and $f_{0}(1710)$ have been found in the literature with the possibilities of other structures like glueball or tetraquark \cite{Frederic}. Whereas, in the Ref.\cite{21,Ju-Jun Xie,Rathaud} the state $f_{2}'(1525)$ and $f_{0}(1710)$ have been predicted as a promising VV molecule. Our calculated mass spectra of the state $f_{2}'(1525)$ and $f_{0}(1710)$ are in agreement with \cite{21,Ju-Jun Xie,Rathaud} and suggest it as VV-molecule. Even more, some states like  $a_{0}(1450)$  and $f_{2}(1810)$ have not confirmed place in the PDG list \cite{21}. \\

In the framework of the present study, we have only incorporated the pion and sigma exchange as SU(2) symmetry. The tensor and spin-orbit coupling may have significant contribution which are very useful for L$\neq$0 as well as for mixed states. To confirm the molecular structures of all these states, one has to study other properties like decay width, possible coupling, etc. Our aim of the present study is to give the attention to all these states which may have a possible molecular internal structure and the predicted mass spectra will provide reference for future study of molecular dimesonic states. We will take this study one step further ahead by including other mesons contributions with possible S-D wave mixing, as our future study.\\

Finally, in this paper, we are able to calculate the S-wave masses and digamma  decay widths of the dimesonic states in the light meson sector. We would like to  improve and implement this model to calculate heavy flavour mesons as well as for higher excited states.  \\
	
\noindent {\bf Acknowledgements}
A. K. Rai would like to thanks Prof. Atsushi Hosaka from Osaka University for the useful discussion during the Hadron Spectroscopy 2013, and also acknowledge the financial support extended by D.S.T., Government  of India  under SERB fast track scheme SR/FTP /PS-152/2012.\\

\end{document}